\documentclass[journal]{IEEEtran}
\usepackage{cite}
\ifCLASSINFOpdf
\else
\fi
\usepackage{amsmath}
\usepackage{mathtools}
\usepackage{xcolor}
\begin{document}
\title{Arbitrary Wave Transformations with Huygens' Metasurfaces through Surface-Wave Optimization}

\author{Vasileios~G.~Ataloglou,~\IEEEmembership{Student Member,~IEEE,}
        and~George~V.~Eleftheriades,~\IEEEmembership{Fellow,~IEEE}% <-this % stops a space
\thanks{The authors are with the Edward S. Rogers Sr. Department of Electrical and Computer Engineering, University of Toronto, Toronto, Canada (e-mail: vasilis.ataloglou@mail.utoronto.ca; gelefth@waves.utoronto.ca).}}% <-this % stops a space

\maketitle

\begin{abstract}
Huygens' metasurfaces have demonstrated the ability to tailor electromagnetic wavefronts with passive low-profile structures. The fundamental constraint enabling passive and ideally lossless solutions is the conservation of the normal real power locally along the metasurface. In this work, we examine the use of auxiliary surface waves to overcome this limitation and design Huygens' metasurfaces for wave transformations with different incident and output power density profiles. The developed method relies on the optimization of a surface-wave distribution that is utilized to redistribute the power at the input side of the metasurface without incurring any reflections. A full design example is presented with a linear patch array along the $H$-plane illuminating a metasurface that produces uniform output fields along the $E$-plane. A high aperture illumination efficiency of $92\%$ is obtained despite the small distance between the source and the metasurface. Moreover, the effects of the evanescent spectrum to the losses and the bandwidth of the structure are discussed.

\end{abstract}


% \fi
\IEEEpeerreviewmaketitle

\section{Introduction}


\IEEEPARstart{M}{etasurfaces} have emerged as a promising platform to shape electromagnetic wavefronts at will. In particular, passive Huygens' metasurfaces (HMSs) are embodied by sub-wavelength scatterers inducing co-located electric and magnetic dipole moments when illuminated by an incident field \cite{Pfeiffer:PRL2013,Selvanayagam:Opt2013}. The electrically small dimensions of the scatterers (also called ``unit cells'') allow the HMS to be homogenized and described by continuous parameteres such as surface impedances/admittances or surface susceptibilities \cite{Achouri:TAP2015,Epstein:TAP2016}. The design process of HMSs usually begins with determining the necessary surface parameters for a desired wave transformation. In a second step, the parameters are discretized and realized by engineering the individual unit cells. Using this approach, numerous wave transformations such as perfect anomalous refraction and beam focusing have been demonstrated \cite{Chen:PRB2018,Lavigne:TAP2018,Chen:TAP2019Lens}.

The fundamental constraint of a wave transformation performed by a passive HMS is that the power density has to be locally conserved. This is the case even for omega-bianisotropic HMSs, which allow to match different wave impedances at the two sides but do not remove the local power conservation requirement \cite{Epstein:TAP2016}. To address this issue, metasurface pairs have been proposed to redistribute the power in the region between the two metasurfaces \cite{Raeker:PRL2019, Ayman:AWPL2018, Ataloglou:TAP2020}. Another concept is the use of a cavity-excited metasurface, where reflections restore power matching; yet, no power is lost, since it is confined within a cavity \cite{Epstein:TAP2017Antennas,Farahabadi:JOSAB2019}. However, both approaches increase the size of the total structure.

Alternatively, surface waves have demonstrated capabilities of redistributing the power density profile without inducing any reflected propagating power. In \cite{Epstein:PRL2016}, it was analytically shown that two counter-propagating surface waves can be utilized for beam splitting of a normally incident plane wave. Conceptually, surface waves decay away from the HMS, so they do not alter radiation in the far-field region. However, when superimposed with the incident fields they create additional power-density terms that fall into the visible spectrum and modify the power-density profile. In recent works, a set of surface waves has also been introduced and optimized for more complex applications, such as beamforming with impenetrable reflective metasurfaces and cloaking \cite{Kwon:PRB2018,Kwon:PRB2018cloacking}. Lastly, Method of Moments approaches taking into account all mutual coupling interactions (propagating or evanescent) between elements have also been used recently for beamforming with metasurfaces \cite{Budhu:TAP2021,Pearson:arxiv2020}.

In this paper, we discuss a design method, first presented in \cite{Ataloglou:APS2020}, for performing wave transformations characterized by different incident and output power-density profiles with a single transmissive HMS exciting surface waves. Additionally, we examine the feasibility of the concept with the design and simulation of a physical HMS performing such a wave transformation. Operating in the transmission mode, results in structures that do not suffer from aperture blockage due to scattering from the source, as happens for example in conventional reflectors. Moreover, by introducing and optimizing a surface-wave distribution, it is possible to ideally transmit all the incident power to the output of the HMS according to a desired output tapering. To validate our formulation, we present an example of obtaining uniform fields from an HMS placed above a linear patch array. Practical considerations, such as power losses and bandwidth are also discussed.

\section{Design method}
\begin{figure}[!t]
\centering
\includegraphics[width=0.8\columnwidth]{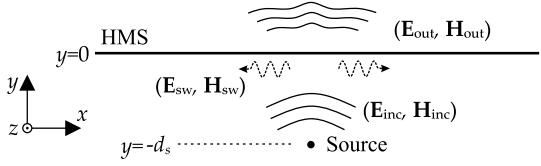}
\caption{Sketch of an HMS exciting surface waves $(\mathbf{E}_\mathrm{sw},\mathbf{H}_\mathrm{sw})$ in order to alleviate the local power mismatch between the incident and the output power densities.}
\label{fig:Fig1}
\end{figure}

In this section, the method for calculating the surface-wave distribution for a desired wave transformation is detailed. The geometry under consideration is sketched in Fig.~\ref{fig:Fig1}. For simplicity we assume uniformity along the $z$-axis and transverse-magnetic (TM) polarized fields. The HMS, placed at the plane $y=0$, has a total width of $L_\mathrm{tot}$ and is illuminated by a source at the plane $y=-d_s$. The incident fields $(E^x_\mathrm{inc},H^z_\mathrm{inc})$ are determined based on the source illumination and the output fields $(E^x_\mathrm{out},H^z_\mathrm{out})$ are specified according to the desired far-field radiation. While the incident and output power density profiles can be different in principle, the total power is set to be equal at the two sides, so that all the incident power is transmitted. The problem is then reduced to determining the surface waves that restore local power conservation.

The unknown surface-wave distribution is introduced as a summation of modulated spatially-shifted sinc functions:
\begin{align} \label{eq:Hsw}
H_\mathrm{sw}^z(x,y=0^-)\!= \!\! \sum_{n=-N}^{n=N}\!\! A_n  \frac{\mathrm{sin}(k_w (x-nL))}{k_w (x-nL)} \mathrm{cos}(k_c(x-nL)),
\end{align}
where $A_n$ are complex unknown weights, $L$ is the distance between two consecutive basis-functions and $k_w, k_c$ are wavenumbers that determine the range of added spectrum and the center spatial frequency, respectively.  Specifically, by taking the Fourier Transform, the spectrum is calculated as
\begin{align} \label{eq:Hsw-FT}
\tilde{H}^z_{\mathrm{sw}}(k_x)\! =\!
\begin{dcases} 
\sum_{n=-N}^{n=N} \frac{\pi A_n}{2k_w}  e^{+j k_x nL} &\! \! , |k_x| \in [k_c-k_w,k_c+k_w]\\
0 &\! \!  ,\mathrm{otherwise}.
\end{dcases}
\end{align}
As seen from Eq.~\eqref{eq:Hsw-FT}, by choosing $k_c-k_w>k$ ($k$ being the free-space wavenumber), it is possible to have a purely evanescent continuous spectrum for the added fields. The magnetic field at the entire input side ($y<0$) is computed as an inverse Fourier Transform (IFT),
\begin{align} 
H^z_{\mathrm{sw}}(x,y)&=\mathrm{IFT} \{ \tilde{H}_{\mathrm{sw}}^z(k_x,y=0^-) e^{+jk_yy} \} \nonumber \\
= \sum_{n=-N}^{n=N} & \frac{A_n}{2 k_w}  \int_{k_c-k_w}^{k_c+k_w} e^{\sqrt{k_x^2-k^2}y} \mathrm{cos}(k_x(x-nL)) dk_x,
\end{align}
where the wavenumber $k_y=-j\sqrt{k_x^2-k^2}$ is chosen such to express an exponential decay away from the HMS in the input region ($y<0$). This guarantees that the far-field will not be perturbed by the presence of the surface-wave distribution. Finally, the electric field corresponding to the surface-wave distribution in Eq.~\eqref{eq:Hsw} is calculated in the $y=0^-$ plane as
\begin{align} \label{eq:Esw}
& \!\! E^x_{\mathrm{sw}}(x,y=0^-)=-\frac{j}{\omega \varepsilon} \frac{ H^z_{\mathrm{sw}}}{\partial y}\Bigg|_{y=0^-} \nonumber \\
& \!\!= \! \sum_{n=-N}^{n=N} \frac{-j \eta A_n}{2  k_w k} \! \int_{k_c-k_w}^{k_c+k_w} \!\!\! \sqrt{k_x^2-k^2} \mathrm{cos}(k_x(x-nL)) dk_x,
\end{align}
where $\eta \approx 120 \pi \ \Omega$ is the impedance of free-space.

Having expressions for both the electric and the magnetic field at the plane $y=0^-$ allows the calculation of the input power density as a function of the weights $A_n$ in the presence of the auxiliary surface waves. In particular, by superimposing the incident and auxiliary fields, the real part of the Poynting vector normal to the HMS is:
\begin{align}\label{eq:Pin}
P_{\mathrm{in},y}(x)=-\frac{1}{2} \mathrm{Re} \Big\{ \big(E^x_{\mathrm{inc}}&+E^x_{\mathrm{sw}}(x,y=0^-)\big) \nonumber \\
&\big(H^z_{\mathrm{inc}}+H^z_{\mathrm{sw}}(x,y=0^-)\big)^* \Big\}.
\end{align}  
Expanding the terms in Eq.~\eqref{eq:Pin}, four terms will appear; the term $E^x_{\mathrm{inc}} H^{z*}_{\mathrm{inc}}$ represents the incident power, while the other three terms can modify the total power-density profile at $y=0^-$ so that it matches the output. To achieve this, the normal power is matched at a number of equidistant points along the boundary which coincide with the center of each unit cell and the centers of the basis functions defined in Eq.~\eqref{eq:Hsw}. As a result of this power-based point-matching process, a nonlinear system of equations is formed in terms of the weights $A_n$ as
\begin{align} \label{eq:G}
\mathbf{G}=\Big[  
P_{\mathrm{in},y}-P_{\mathrm{out},y}\Big]_{x=nL} = 0, \ \ -N \leq n \leq N,
\end{align}
where $P_{\mathrm{out},y}=-\mathrm{Re}\{E^{x}_\mathrm{out}H^{z*}_\mathrm{out}  \}/2$ is the output power density based on the desired fields and $\mathbf{G}$ is a column vector with each row corresponding to the power mismatch at a different point across the HMS. Since the system of equations in \eqref{eq:G} cannot be solved analytically, gradient-descent optimization is employed to minimize the total power mismatch, expressed as
\begin{align}
F=\frac{1}{2} \mathbf{G}^T \mathbf{G},
\end{align}
with $T$ denoting a conjugate operation.

The optimization has been proven robust for a number of wave transformations including uniform or tapered apertures (e.g. for a Taylor or Chebyshev radiation pattern) and for various sources (e.g. linesource, patch array along $z$-axis). To this direction, the choice of basis-functions that decay away from the center is beneficial, as every variable $A_n$ affects mostly the fields around $x=nL$. After a solution has been obtained from the optimization process, the HMS parameters are calculated based on closed-form expressions involving the total tangential fields at the two sides \cite{Epstein:TAP2016}. In particular, the metasurface is characterized by a surface electric impedance $Z_\mathrm{se}$, a surface magnetic admittance $Y_\mathrm{sm}$ and a magnetoelectric coupling coefficient $K_\mathrm{em}$. Realizing these parameters results in an HMS that transforms the total input fields to the desired output fields. To this purpose, stacked-layer unit cells are used, as presented in Sec.~III(b).

\section{Uniform output aperture design}
In this section, we present a design example of an HMS exciting surface waves using the method described in Sec.~II. The source is a linear patch array and uniform fields are required at the output.
\subsection{Geometry and source design}
\begin{figure}[!t]
\centering
\includegraphics[width=0.9\columnwidth]{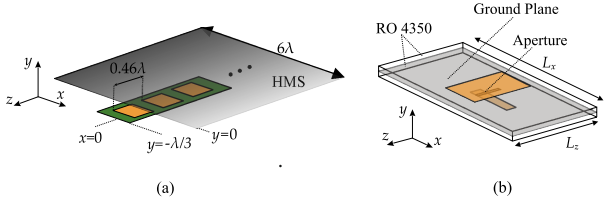}
\caption{(a) Linear patch array along $z$ illuminates an HMS. Surface waves need to be excited along the $x$ direction to achieve uniform output tapering. (b) Geometry of the aperture-fed patch element.}
\label{fig:Fig2}
\end{figure}

The geometry of the structure is depicted in Fig.~\ref{fig:Fig2}(a). The source is a linear array of patch antennas operating at $10 \ \mathrm{GHz}$, placed at a distance $d_s=\lambda/3$ away from the HMS. Although the patch array and the HMS are assumed to be infinite along the $z$-axis for the purposes of simulations, the structure can be truncated to $16$ elements since these are sufficient to maintain uniformity along the $z$-direction for the majority of the HMS area. The HMS has a width of $L_\mathrm{tot}=6\lambda$ along the $x$-axis and it is required that the output fields are uniform along the aperture, so that the directivity is maximized.

A single patch element is shown in Fig.~\ref{fig:Fig2}(b). The patch has dimensions $10 \ \mathrm{mm} \times 6.51 \ \mathrm{mm}$ and is aperture-fed by a rectangular slot of dimensions $4 \ \mathrm{mm} \times 0.5 \ \mathrm{mm}$. Both substrates are Rogers $\mathrm{RO}4350$ ($\epsilon_r=3.66 , tan \delta=0.004$) of thickness $t=0.762 \ \mathrm{mm}$, while the width of the ground plane of the patch along the $x$-direction is $L_x=\lambda$. The choice of aperture feeding for the patches was preferred both due to the slightly higher bandwidth and the better isolation between the transmission-line network and the HMS. The distance between the elements is adjusted to $L_z\approx 0.46 \lambda$ to match exactly the total width of $3$ unit cells along the $z$-axis. This is not necessary, since the fields of the patch array are nearly uniform along $z$, but it is convenient for efficient full-wave simulations.

\subsection{Optimization Algorithm}
\begin{figure}[!t]
\centering
\includegraphics[width=0.9\columnwidth]{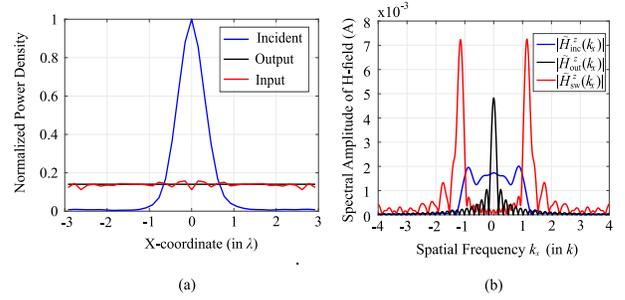}
\caption{(a) Power density of incident (blue curve) and output (black curve) waves. Power mismatch is restored when adding the optimized surface-wave distribution (red curve). (b) Spectrum of incident (blue curve), output (black curve) and auxiliary (red curve) magnetic field $H^z$.}
\label{fig:Fig3}
\end{figure}

The incident power-density profile $P_\mathrm{inc}(x)$ is calculated from a simulation without the HMS. As observed in Fig.~\ref{fig:Fig3}(a), the incident power is very concentrated around $x=0$ compared to the power of a uniform aperture. Hence, surface waves need to be excited and carry the surplus power from the center of the HMS towards the edges. Regarding the field distribution at Eq.~\eqref{eq:Hsw}, the wavenumbers are set to $k_c=2.1k$ and $k_w=1.1k$ resulting in an added spectrum at $[k,3.2k]$. Moreover, the HMS is discretized to $2N+1=39$ unit cells with a width $L \approx \lambda/6.5$. Regarding the discretization, the largest possible width for each unit cell was preferred to facilitate their realization, while still obeying the sampling theorem for the maximum added spatial frequency $L < 2\pi/2k_\mathrm{max}=\lambda/6.4$. 

The input power, after the optimization has converged, closely matches the output power density as evident from Fig.~\ref{fig:Fig3}(a). It is also worth discussing the spectrum of the incident fields, the added surface-wave distribution as determined from the optimization and the total input fields. Figure~\ref{fig:Fig3}(b) reveals that the added spectrum is mainly confined in the $[k,3.2k]$ range. Some leakage into the visible spectrum can be observed due to the finiteness of the HMS; yet, the visible spectrum of the added fields is much weaker than that of the incident wave and no noticeable reflections are expected.

\subsection{Unit Cell Realization}
After determining the necessary surface waves, the ideal parameters $(Z_\mathrm{se}, Y_\mathrm{sm}, K_\mathrm{em})$ are defined at each unit cell. As showed in \cite{Epstein:TAP2016}, it is possible to reformulate the boundary conditions in terms of a $[Z]$ matrix of a two-port microwave network. Then, the required response can be obtained by forming a stacked-layer unit cell comprising parallel admittance sheets separated by substrates acting as transmission-line sections. Naturally, three admittance layers are sufficient to match the three independent HMS parameters $(Z_\mathrm{se}, Y_\mathrm{sm}, K_\mathrm{em})$. However, a fourth layer can be used as an additional degree of freedom to reduce losses by avoiding highly-resonant layers.

\begin{figure}[!t]
\centering
\includegraphics[width=0.9\columnwidth]{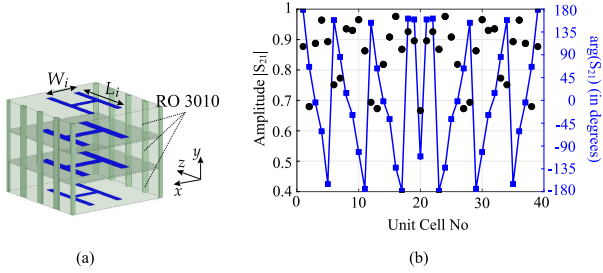}
\caption{(a) Unit Cell structure. (b) Generalized $S_{21}$ for each unit cell: the simulated amplitude $|S_{21}|$ (black dots) is limited by losses, while the phase $\mathrm{S_{21}}$ (blue squares) matches the desired values (continuous blue curve).}
\label{fig:Fig4}
\end{figure}

The unit cells for our example are shown in Fig.~\ref{fig:Fig4}(a). Specifically, four copper layers are used separated by Rogers $\mathrm{RO}3010$ substrates of thickness $t_s=1.27 \ \mathrm{mm}$ and bonded together with Rogers bondply $\mathrm{RO}2929$ of thickness $t_b=0.051 \ \mathrm{mm}$. The width of the copper traces, in the majority of cases, is $s_i=0.4 \ \mathrm{mm}$ for the two center layers and $s_i=0.25 \ \mathrm{mm}$ for the two outer layers. This choice reduces losses for the middle layers that generally have higher admittance values and allows a higher range of realizable admittance values for the outer layers. The width of the dogbones was adjusted to a maximum $W_i=3.7 \ \mathrm{mm}$, while the length was set to $L_i=3 \ \mathrm{mm}$ with few exceptions. Coupling between adjacent unit cells is not considered when designing them individually with periodic boundary conditions. To eliminate these coupling effects, copper vias are used as baffles between the layers \cite{Xu:TAP2019}. While vias isolate each unit cell from the others, they do not inhibit the propagation of the added surface waves, since these extend in the region before the HMS ($y<0$). Vias act also as parallel perfect electric conductor (PEC) plates that enforce transverse electromagnetic (TEM) propagation within the unit cell by prohibiting propagation of higher-order modes \cite{Xu:TAP2019}. The vias' diameter is $D_v=0.33 \ \mathrm{mm}$ and the distance between their centers is $r_v=1.15 \ \mathrm{mm}$ (see Fig.~\ref{fig:Fig4}(a)). An alternative way to account for coupling between different unit cells would be to treat layers as individual boundaries, as presented in \cite{Kwon:AWPL2021}.

The unit cells can be assessed in terms of a generalized scattering matrix $S$, where the port impedances are set to the wave impedance $\mathrm{E}^x/\mathrm{H}^z$ of the total input/output waves at each unit cell. Given that local power is conserved, the ideal lossless case would then correspond to $|S_{21}|=1$ and appropriate phase. As noticed in Fig.~\ref{fig:Fig4}(b), losses reduce the obtained $|S_{21}|$ (black dots), but the phases of $S_{21}$ (blue squares) match satisfactorily with the desired values (blue curve).

\subsection{Full-wave Simulations}
\begin{figure}[!t]
\centering
\includegraphics[width=\columnwidth]{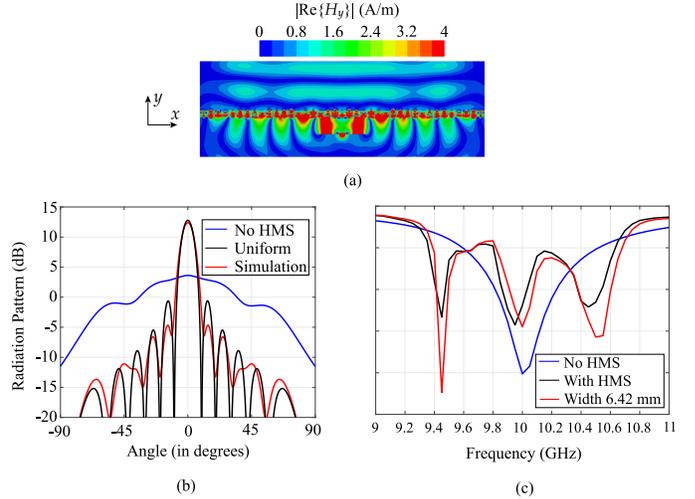}
\vspace{-1em}
\caption{Full-wave simulation results: (a) Magnetic field $|\mathrm{Re} \{H^z\}|$. (b) Radiation pattern of the simulated transmitted fields (red curve), a $6\lambda$ uniform aperture (black curve) and the patch array (blue curve) in the $xy$ plane. (c) Reflection coefficient for the patch array alone (blue curve), with the HMS (red curve) and with the HMS and a patch width of $6.42 \ \mathrm{mm}$ (black curve).}
\label{fig:Fig5}
\end{figure}

The structure is simulated with ANSYS HFSS and the magnetic field distribution $|\mathrm{Re} \{H^z\}|$ is plotted in Fig.~\ref{fig:Fig5}(a). As observed, the output field indeed exhibits close to uniform amplitude and phase with some amplitude tapering observed only towards the edges of the HMS. Strong evanescent waves can also be spotted at the input side close to the HMS. Naturally, this evanescent distribution is higher at the center of the metasurface ($x=0$), since the mismatch between the incident and output power is higher at this part of the HMS. It is noted that the structure in Fig.~\ref{fig:Fig5}(a) is surrounded by absorbing boundary conditions and there is no metallic cavity involved, but the power is redistributed through the auxiliary surface waves.

Figure~\ref{fig:Fig5}(b) shows the obtained $2$-D (in the $x-y$ plane) radiation pattern of the transmitted fields in comparison with the one corresponding to a uniform aperture distribution. The satisfactory agreement in terms of maximum directivity and gradually decaying side lobes reveals good accuracy of the performed transformation. In particular, the obtained directivity is $12.39 \ \mathrm{dB}$ compared to $ 12.73 \ \mathrm{dB}$ in the uniform case, resulting in an aperture illumination efficiency of $92 \%$ (defined as the obtained directivity divided by the directivity of a uniform aperture). The small deviations are attributed to the non-uniform losses of the unit cells and to the perturbations introduced from the near-field coupling between the source and the metasurface. It should be highlighted that an output aperture with uniform phase but amplitude tapering according to the incident fields would have a maximum aperture illumination efficiency of only $44 \%$. That would be the case if an omega-bianisotropic HMS was designed but without exciting any surface waves to redistribute the incident power.

Lastly, we characterize the HMS in terms of losses and bandwidth. The power transmission efficiency, defined as the transmitted power at the output of the HMS divided by the incident power, is estimated to be $67 \%$, limited primarily by dielectric and copper losses. The slightly increased power losses compared to other HMS designs are associated with the strong evanescent fields that interact with the metasurface. With regard to the bandwidth, a lumped port is placed in the transmission line that feeds the patch (with periodic boundary conditions along $z$). This allows to calculate $S_{11}$ both with and without the metasurface. As seen in Fig.~\ref{fig:Fig5}(c), the HMS shifts slightly the center frequency and reduces the fractional bandwidth ($S_\mathrm{11}<-10 \ \mathrm{dB}$) to $1.7 \%$ from $4.1 \%$. Within this range the directivity is maintained above $11.45 \ \mathrm{dB}$. Moreover, the shift in the center frequency can be corrected by tuning the width of the patch to $6.42 \ \mathrm{mm}$ (from $6.51 \ \mathrm{mm}$ initially). The relatively narrowband behavior is attributed to the evanescent fields, the resonant nature of some unit cells and the narrowband matching of the patch array itself.

\section{Conclusion}

A method to design metasurfaces has been proposed for the case of different incident and output power density profiles. The method relies on the excitation of a surface-wave distribution that restores local power matching without inducing reflections. A case study has been presented with a metasurface being illuminated by a uniform linear patch array extending along the $H$-plane. The incident fields along the $E$-plane are converted to uniform output fields with the use of auxiliary surface waves, passively excited by the metasurface. The metasurface is implemented with four-layer unit cells that are designed individually based on the desired response. The importance of surface waves to realize a highly-directive beam is discussed, especially if the source is placed close to the HMS. In the example examined, more than a twofold increase was obtained in terms of aperture illumination efficiency ($92 \%$ versus $44\%$) when comparing with an ideal metasurface acting only on the phase of the transmitted fields. The method can also be used to achieve arbitrary far-field radiation patterns by specifying the corresponding output aperture fields.

% or


\appendices

\section*{Acknowledgment}
Financial support from the Natural Sciences and Engineering Research Council of Canada is gratefully acknowledged. V.~A. also acknowledges the support of the Ministry of Advanced Education and Skills Development of Ontario, the Electrical and Computer Engineering Department of the Univ. of Toronto and the Alexander S. Onassis Foundation.

\ifCLASSOPTIONcaptionsoff
  \newpage
\fi

\end{document}